\documentclass{article}

\usepackage[nonatbib, preprint]{neurips_2024}

\usepackage[utf8]{inputenc} 
\usepackage[T1]{fontenc}    
\usepackage{hyperref}       
\usepackage{url}            
\usepackage{booktabs}       
\usepackage{amsfonts}       
\usepackage{nicefrac}       
\usepackage{microtype}      
\usepackage{xcolor}         

\usepackage[doi=false,isbn=false,eprint=false,sorting=none,minnames=3]{biblatex}
\usepackage{graphicx}
\usepackage{amsmath}

\title{SALMONN-omni: A Codec-free LLM for Full-duplex Speech Understanding and Generation}

\author{%
  Wenyi Yu$^{1,2}$\thanks{These authors contributed equally to this work.} \quad Siyin Wang$^{1,2*}$ \quad Xiaoyu Yang$^{3}$ \quad Xianzhao Chen$^{2}$ \quad Xiaohai Tian$^{2}$\\ \textbf{Jun Zhang}$^{2}$ \quad \textbf{Guangzhi Sun}$^{3}$ \quad \textbf{Lu Lu}$^{2}$ \quad \textbf{Yuxuan Wang}$^{2}$ \quad \textbf{Chao Zhang}$^{1}$\thanks{Corresponding author.} \\
  $^1$Tsinghua University \quad $^2$ByteDance \quad $^3$University of Cambridge\\
  \texttt{\{ywy22,wangsiyi23\}@mails.tsinghua.edu.cn} \quad
  \texttt{cz277@tsinghua.edu.cn}
}

\begin{document}

\maketitle

\begin{abstract}
Full-duplex multimodal large language models (LLMs) provide a unified framework for addressing diverse speech understanding and generation tasks, enabling more natural and seamless human-machine conversations. Unlike traditional modularised conversational AI systems, which separate speech recognition, understanding, and text-to-speech generation into distinct components, multimodal LLMs operate as single end-to-end models. This streamlined design eliminates error propagation across components and fully leverages the rich non-verbal information embedded in input speech signals. We introduce SALMONN-omni, a codec-free, full-duplex speech understanding and generation model capable of simultaneously listening to its own generated speech and background sounds while speaking. To support this capability, we propose a novel duplex spoken dialogue framework incorporating a ``thinking'' mechanism that facilitates asynchronous text and speech generation relying on embeddings instead of codecs (quantized speech and audio tokens). Experimental results demonstrate SALMONN-omni's versatility across a broad range of streaming speech tasks, including speech recognition, speech enhancement, and spoken question answering. Additionally, SALMONN-omni excels at managing turn-taking, barge-in, and echo cancellation scenarios, establishing its potential as a robust prototype for full-duplex conversational AI systems. To the best of our knowledge, SALMONN-omni is the first codec-free model of its kind. A full technical report along with model checkpoints will be released soon.
\end{abstract}

\section{Introduction}

Large language models (LLMs) \cite{gpt3,llama,jiang2024mixtral} have established a new approach to problem-solving and task execution through natural conversations. Speech, being a fundamental form of human communication, acts as an intuitive and effective means for interactions between humans and LLMs. As a result, there is a growing research emphasis on enhancing the spoken input and output capabilities of LLMs. Some recent studies have focused on equipping LLMs with a comprehensive understanding of speech and audio \cite{tang2024salmonn,ltu,Qwen2-Audio}, while other research has explored utilizing LLMs’ advanced language understanding abilities to develop more sophisticated speech generation and processing methods \cite{llm4tts_1,dubey2024llama}.

Although half-duplex or turn-based speech LLMs \cite{zhang-etal-2023-speechgpt,nguyen2024spirit,fang2024llama,xie2024mini} can function as conversational AI systems, human conversations are inherently \textit{full-duplex}, characterized by simultaneous listening, thinking, and speaking. This dynamic is reflected in natural conversational behaviours such as frequent turn-taking, backchanneling, overlapping speech, and barge-in. These interactive and flexible patterns have sparked increasing interest in developing full-duplex speech LLMs to enhance the fluidity and naturalness of human-computer interactions.
Traditional modularised systems \cite{young2013pomdp,vinyals2015icml,serban2016aaai} typically follow a multi-stage pipeline: an automatic speech recognition (ASR) system transcribes the user’s speech into text, which is then processed by a task-oriented dialogue (ToD) system \cite{young2013pomdp,wen-etal-2017-network,NEURIPS2020_e9462095} or chatbot \cite{vinyals2015icml,serban2016aaai,see-manning-2021-understanding,zhang-etal-2024-beyond} (often powered by an LLM) to generate a text response. This response is subsequently converted into speech by a text-to-speech (TTS) synthesis system. Additional modules for turn-taking prediction \cite{chang2022turn}, backchanneling \cite{wang2024turn}, device-directed speech detection \cite{chang2022streaming}, barge-in handling \cite{bekal2022contextual}, and echo cancellation \cite{o2021conformer} are often integrated into either the ASR or the natural language understanding components of the ToD system. While these modularised systems support basic human-machine dialogues (typically turn-based and involving a single speaker), their pipeline architecture introduces significant complexity and is prone to error accumulation across subsystems.
Recent advancements in text-based LLMs and the multimodal speech and audio understanding capabilities of LLMs offer a compelling opportunity to develop fully end-to-end, full-duplex systems for natural spoken dialogues. These systems promise to deliver more human-like conversational and speech capabilities with streamlined architectures, paving the way for the next generation of intelligent personal assistants.

Recently, the announcement and release of GPT-4omni's (GPT-4o) advanced speech capabilities have garnered significant global interest in the development of LLM-based, low-latency, emotionally expressive and possibly full-duplex human-machine speech interaction technologies. However, as GPT-4o is a commercial product, the underlying technologies remain undisclosed.
Several codec-based models have been proposed to achieve full-duplex capability in an end-to-end manner \cite{ma2024language,moshi,syncllm}. After tokenizing speech into discrete tokens and modelling both auditory and textual tokens in a single Transformer \cite{vaswani2017attention} model, Moshi \cite{moshi} utilizes parallel streams to model output speech and input speech simultaneously, while SyncLLM \cite{syncllm} proposes modelling tokens from input stream and output stream in an interleaved manner.

This paper presents SALMONN-omni, a speech understanding and generation model built on a novel codec-free full-duplex spoken dialogue framework. By seamlessly integrating a streaming speech encoder, a large language model (LLM), and a streaming speech synthesizer with cross-attention layers into a unified end-to-end architecture, SALMONN-omni processes both input and output speech simultaneously and in real-time. A periodic synchronization mechanism is employed to provide the model with a ``time'' concept, ensuring effective alignment and synchronization of auditory and textual modalities.
In full-duplex conversations, the model continuously processes user speech and environmental sounds, even while speaking. To address this, a novel ``thinking'' mechanism is introduced, featuring two special state transition tokens that enable the model to handle dynamic scenarios such as turn-taking, barge-in, and echo cancellation \textit{etc}.
Experimental results highlight SALMONN-omni's versatility as a unified solution for a wide range of speech tasks, including speech recognition, synthesis, enhancement, dereverberation, target speaker extraction, and spoken question answering. Furthermore, SALMONN-omni demonstrates effective handling of turn-taking and barge-in in simulated experiments, showcasing its potential as a prototype for enabling natural, seamless human-machine conversations.
To the best of our knowledge, SALMONN-omni is the first model to achieve full-duplex speech understanding and generation without using any speech codec.

\begin{figure}[ht]
    \centering
    \includegraphics[width=\linewidth]{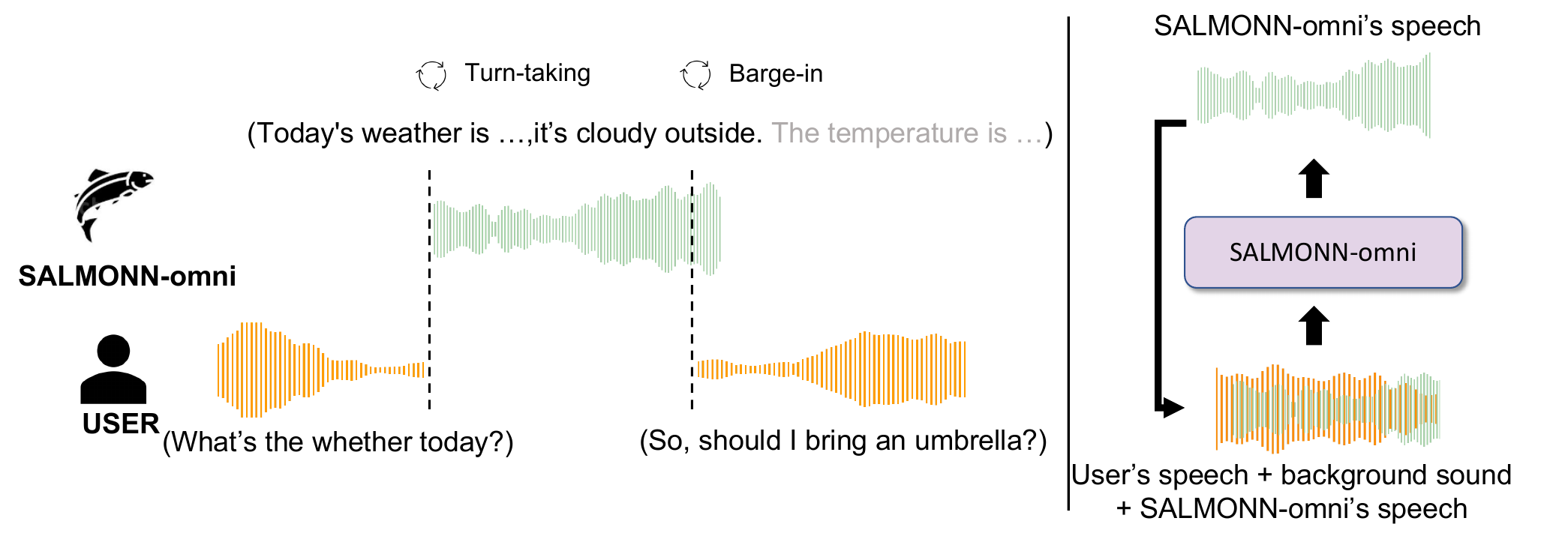}
    \caption{SALMONN-omni is a codec-free, full-duplex end-to-end conversational AI model capable of managing dynamic dialogue interactions such as turn-taking and context-dependent barge-in in human-machine conversations.}
    \label{fig:1}
\end{figure}

\section{Methodology}

Four challenges need to be solved when implementing a full duplex conversational AI model without using codecs.
\begin{itemize}
    \item \textbf{The model needs to support streaming speech input and output.} By integrating an LLM with a streaming speech encoder and a streaming speech synthesizer using embeddings instead of text, SALMONN-omni functions as an end-to-end model, enabling seamless interaction with users through both verbal and non-verbal (paralinguistic) features. 
    \item 
    \textbf{The model should be able to deal with two streams from both the input and the output sides simultaneously.} SALMONN-omni achieves it by utilizing the speech encoder to process the input stream, the LLM to process the output stream and connecting these two components with cross-attention layers.
    \item \textbf{The model should have the idea about ``time'' so that the auditory and textual modalities can be aligned and synchronized.} A periodic synchronization mechanism is introduced for SALMONN-omni. Within each time block, the model processes a fixed duration of input speech and generates a fixed number of textual embeddings.
    \item \textbf{The model should be able to handle complex dynamics in natural conversations such as turn-taking, barge-in, overlapping speech and backchanneling.} A novel ``thinking'' mechanism is proposed so that SALMONN-omni can switch between speaking and non-speaking states flexibly while listening to the input side at all times.
\end{itemize}

\begin{figure}[ht]
    \centering
    \includegraphics[width=\linewidth]{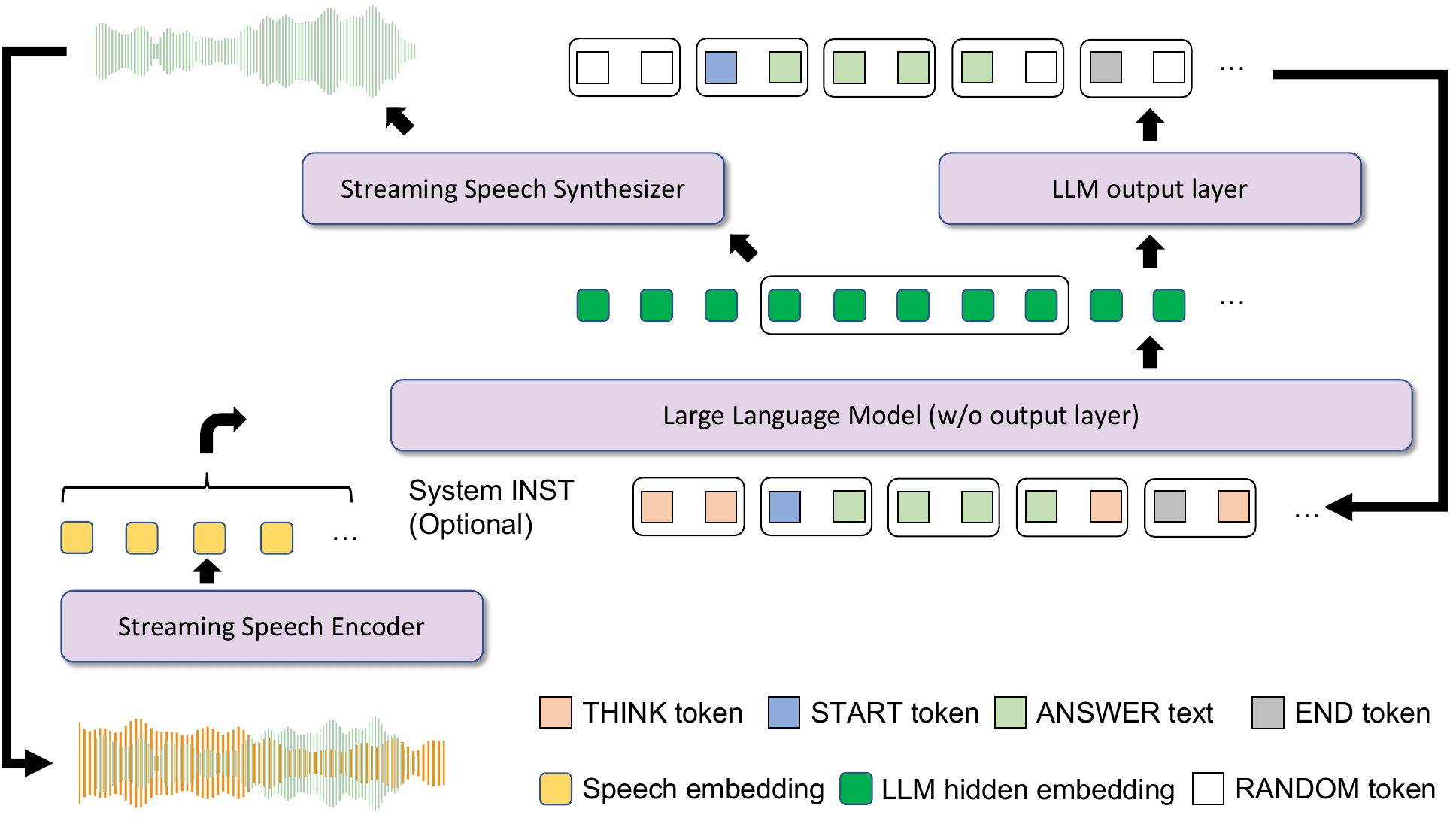}
    \caption{SALMONN-omni is an end-to-end model that integrates a streaming speech encoder, an LLM, and a streaming speech synthesizer, all interconnected through embeddings. It features a novel codec-free, full-duplex spoken dialogue framework enhanced by a ``thinking'' mechanism.}
    \label{fig:2}
\end{figure}

Specifically, SALMONN-omni has speaking and non-speaking states and two special tokens \texttt{<start\_speak>} and \texttt{<end\_speak>} are utilized for transition between these two states. After the LLM generating \texttt{<start\_speak>}, SALMONN-omni switches to the speaking state. When the model completes its answer or is interrupted by the input stream, the LLM generates \texttt{<end\_speak>} and SALMONN-omni transists to the non-speaking state. As shown in Figure \ref{fig:2}, the conversation is cut into a series of time blocks. In block $i$, the streaming speech encoder extracts auditory embeddings from input speech with a fixed duration of $\Delta t$ seconds. Then the LLM generates $n$ word embeddings conditioned on auditory embeddings extracted from block $0$ to block $i$. If the model is in a speaking state, the word embeddings are sent to the streaming speech synthesizer to generate a spoken response with the same duration of $\Delta t$ seconds. There is no explicit text between each of the two components, and the whole model is trained in an end-to-end manner.

To maintain the framework's consistency and simplicity, and to help the LLM determine when to perform state transitions, the LLM is required to decode $n$ tokens within each time block. However, two situations arise where collecting ground truth labels during training becomes challenging:
\textbf{1)} The first occurs in the non-speaking state, where the tokens preceding the \texttt{<start\_speak>} marker are difficult to determine. \textbf{2)} The second arises in the speaking state, when the LLM has completed generating textual embeddings, but the speech synthesizer is still generating and playing audio of the answer. In this case, the LLM must generate tokens to remain informed by the input stream. A straightforward solution might involve introducing a special placeholder token, but the frequent repetition of the same token in the training data risks collapsing the model's output distribution, leading to a significant performance drop.
To address this, the ``thinking'' strategy introduces the special \texttt{<think>} token, which is only used as input in these situations. However, the outputs are not explicitly forced to include this or any other specific tokens. The only constraint is that the outputs must not include \texttt{<start\_speak>} or \texttt{<end\_speak>}. 
This design mirrors the human thought process during conversations, where internal thoughts may differ from spoken words and are not explicitly conveyed. 
Moreover, the mechanism is easy to implement by setting the output labels to either \texttt{<start\_speak>} or \texttt{<end\_speak>}, depending on the state, and applying a negative coefficient $\lambda_\text{think}$ to the loss function. Specifically, if SALMONN-omni is in a non-speaking state, the label is \texttt{<start\_speak>}; otherwise, it is \texttt{<end\_speak>}.

Finally, SALMONN-omni is trained with loss function expressed by Eqn.~\eqref{equ:1},
\begin{equation}
    \mathcal{L}=\lambda_\text{text}\mathcal{L}_\text{text}+\lambda_\text{speech}\mathcal{L}_\text{speech}+\lambda_\text{think}\mathcal{L}_\text{think}, \label{equ:1}
\end{equation}
where $\lambda_\text{text}$, $\lambda_\text{speech}$ and $\mathcal{L}_\text{text}$, $\mathcal{L}_\text{speech}$ are the weights and losses for text generated by the LLM and the final speech response. $\mathcal{L}_\text{think}$ is the loss for ``thinking'' tokens and $\lambda_\text{think}<0$.

\section{Case Study}

SALMONN-omni is trained with multiple tasks such as streaming speech recognition, speech enhancement, spoken question answering and \textit{etc.} Moreover, synthetic data is used for training SALMONN-omni to learn to handle turn-taking and barge-in in natural conversations. The data used for speech recognition include 60k hours of LibriHeavy \cite{kang2024libriheavy} and 10k hours of GigaSpeech \cite{GigaSpeech2021}, which are also the sources of synthetic data for speech enhancement, turn-taking and barge-in.

Below are some cases for demonstrating the capabilities of SALMONN-omni on various streaming speech tasks and natural conversations.

\begin{figure}[h]
    \centering
    \includegraphics[width=\linewidth]{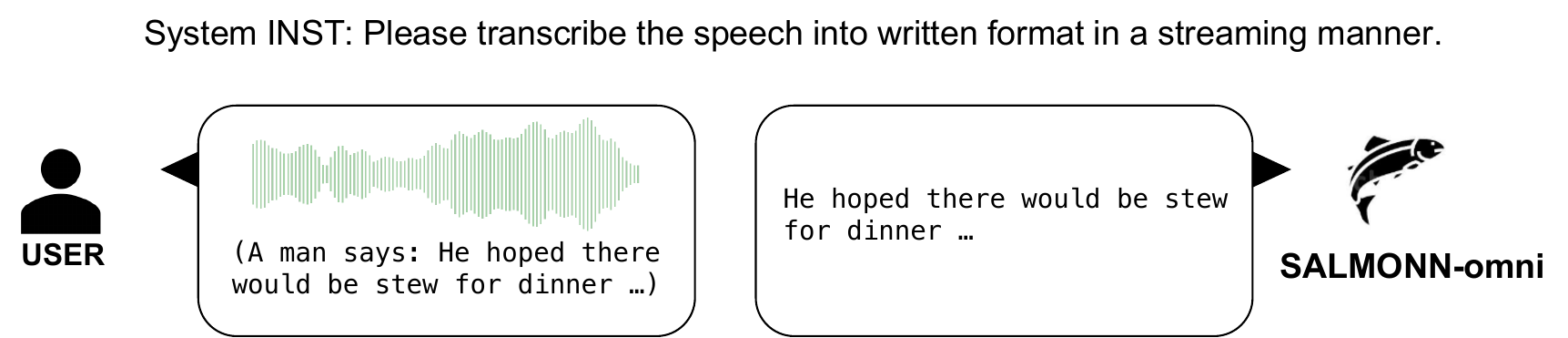}
    \caption{Streaming speech recognition: Without using the speech synthesizer, SALMONN-omni can convert the input speech to text in a streaming way.}
    \label{fig:enter-label}
\end{figure}

\newpage
\begin{figure}[h]
    \centering
    \includegraphics[width=\linewidth]{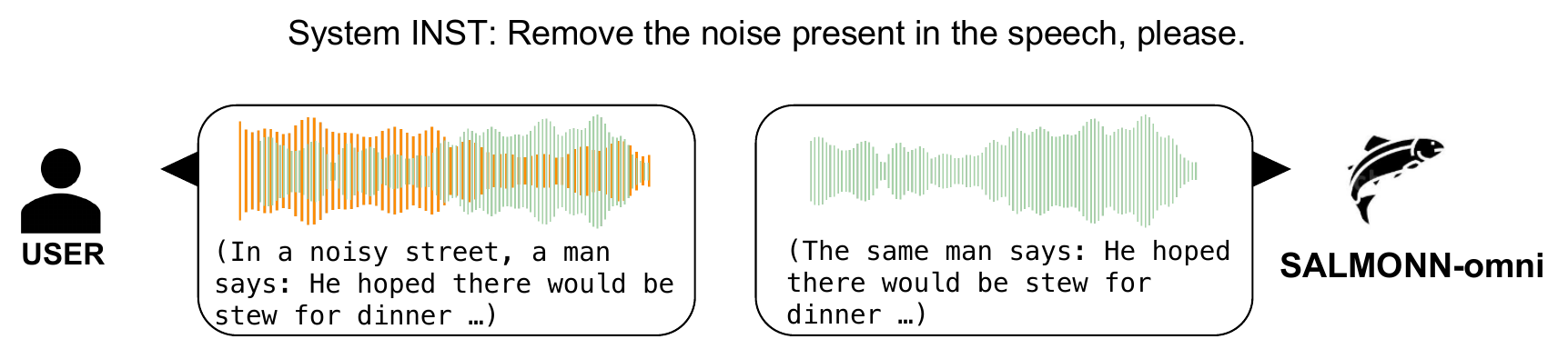}
    \caption{Speech enhancement: SALMONN-omni improves speech quality by denoising and dereverberation. It listens to the noisy speech and then re-speaks the speech content with enhanced clarity.}
    \label{fig:enter-label}
\end{figure}

\begin{figure}[h]
    \centering
    \includegraphics[width=\linewidth]{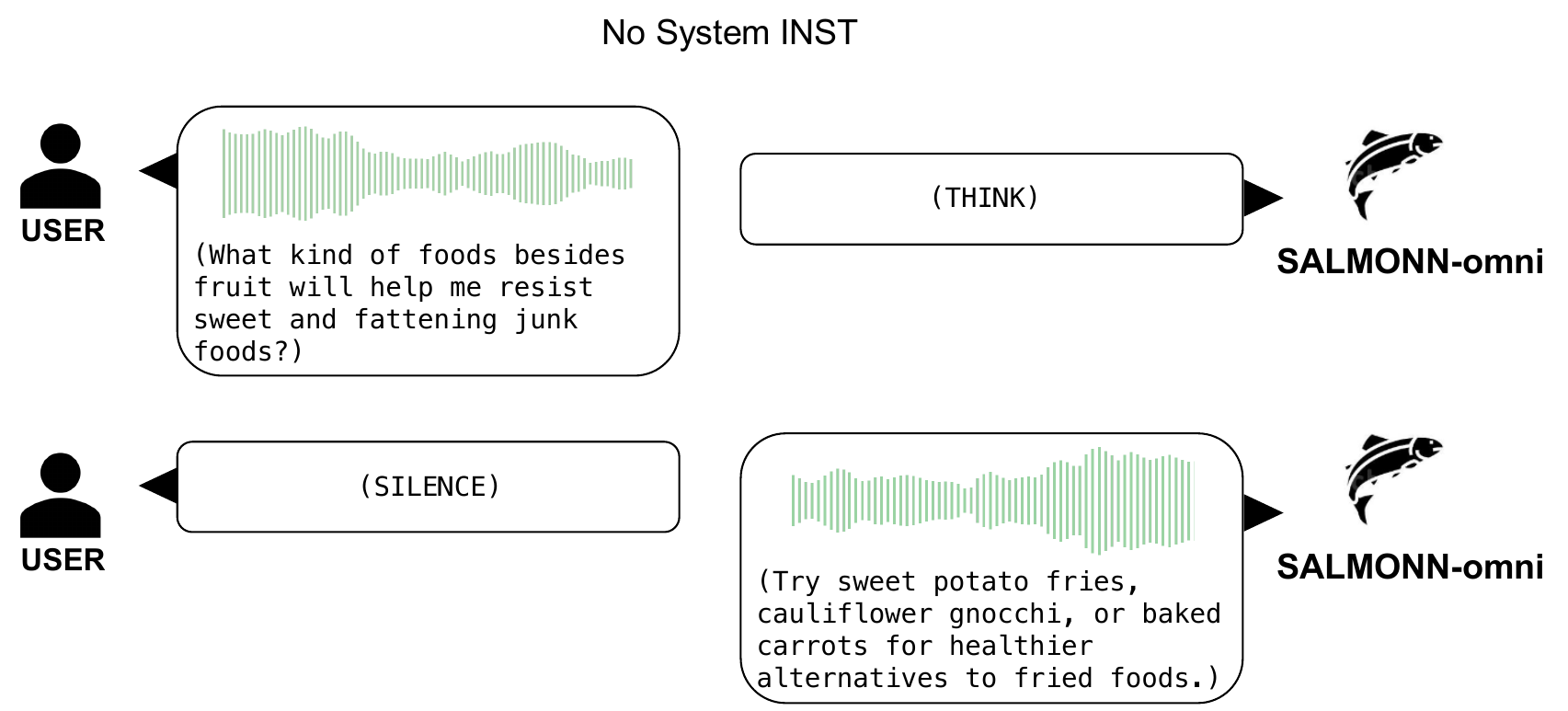}
    \caption{Spoken question answering: SALMONN-omni can handle turn-taking in spoken question-answering scenarios.}
    \label{fig:enter-label}
\end{figure}

\newpage
\begin{figure}[h]
    \centering
    \includegraphics[width=\linewidth]{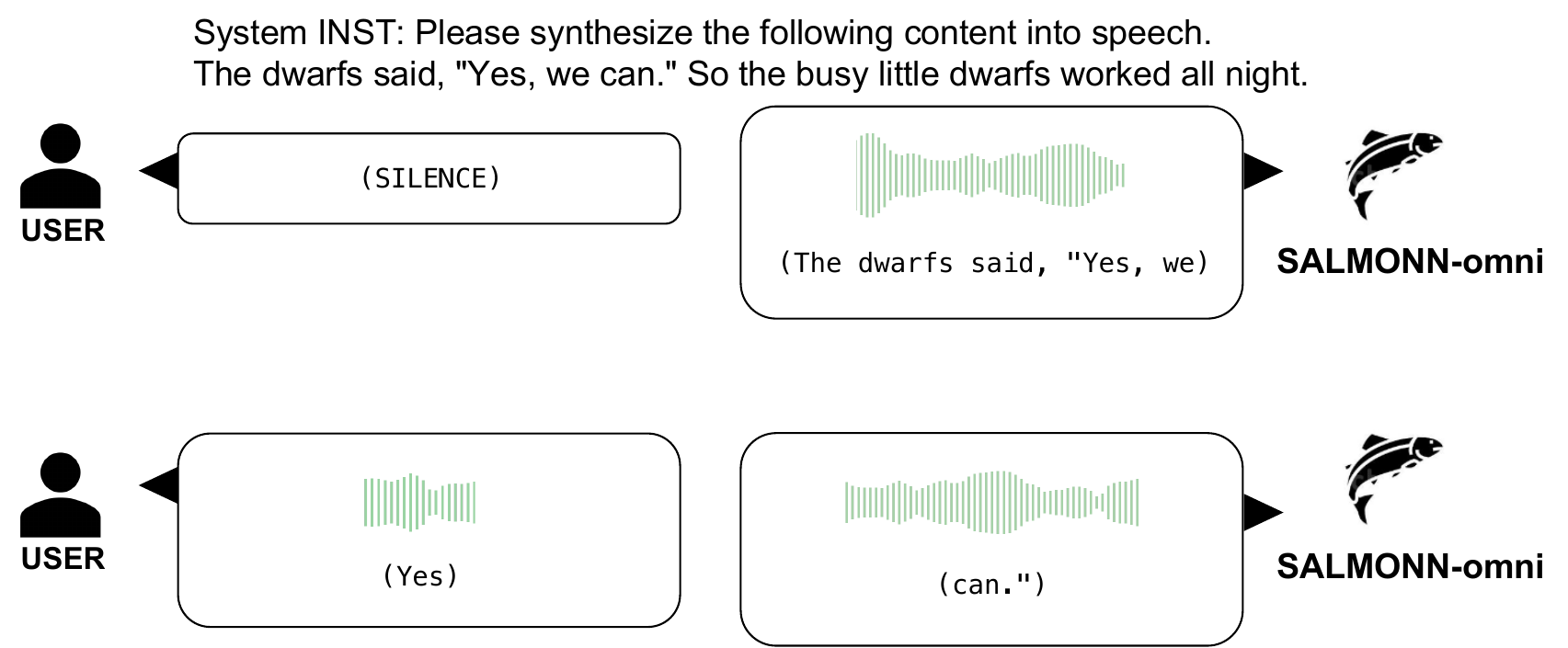}
    \caption{Context-independent barge-in: In this experiment, SALMONN-omni is trained to stop generating speech immediately when the user says ``Yes,'' but continues speaking when it hears other words or does not hear any words.}
    \label{fig:enter-label}
\end{figure}

\begin{figure}[h]
    \centering
    \includegraphics[width=\linewidth]{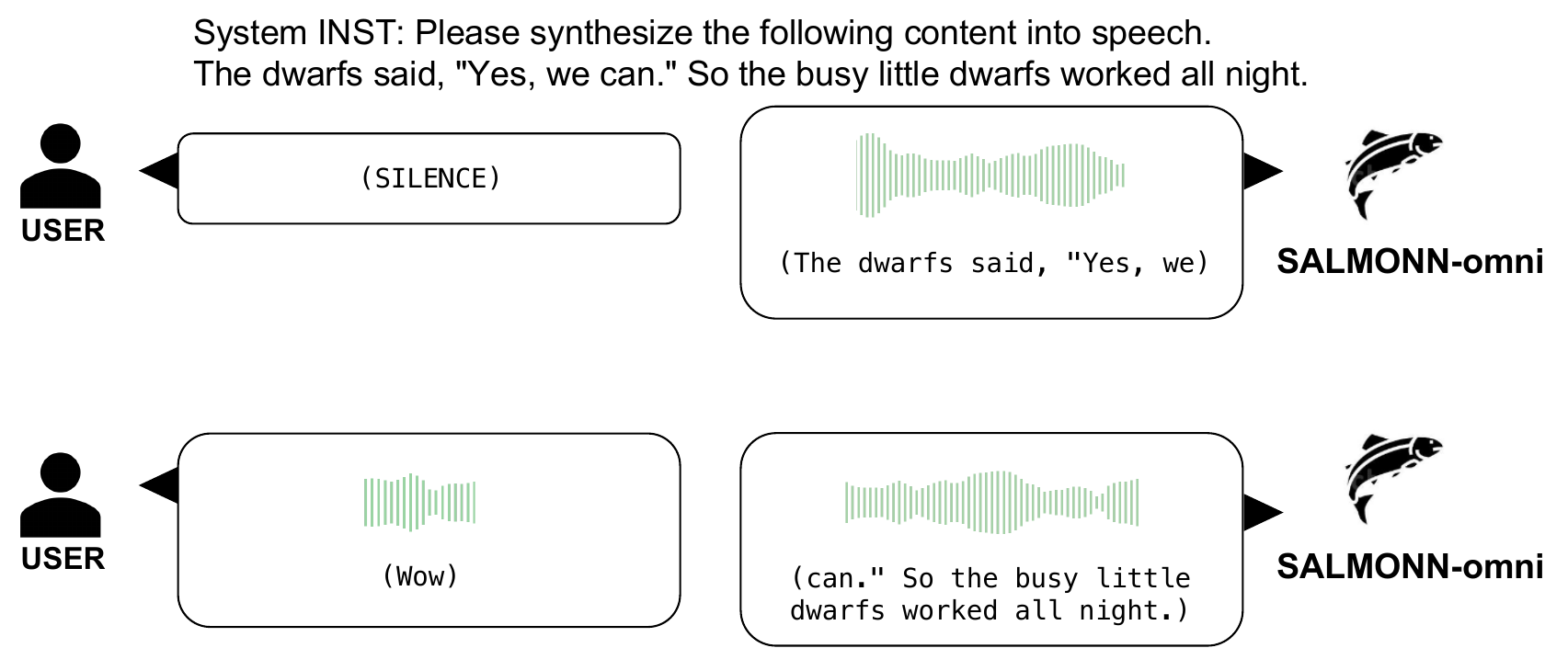}
    \caption{Context-independent barge-in with echo cancellation: In this experiment, SALMONN-omni is trained to immediately stop generating speech upon detecting the user saying "Yes," while continuing uninterrupted when hearing other words or no speech at all. Additionally, SALMONN-omni remains unaffected by its own speech, demonstrating its robust echo cancellation capability.}
    \label{fig:enter-label}
\end{figure}

\newpage
\begin{figure}[h]
    \centering
    \includegraphics[width=\linewidth]{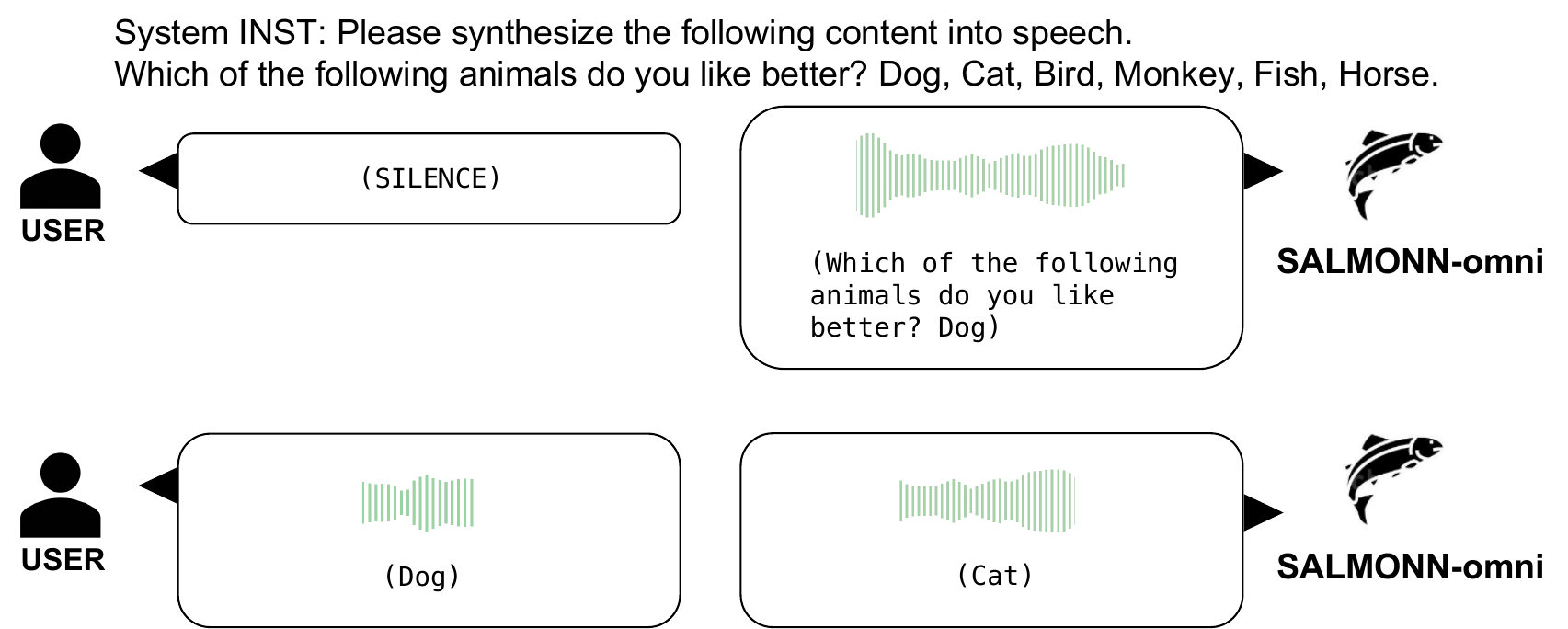}
    \caption{Context-dependent barge-in: When the user is quick to buzz in with the response, SALMONN-omni can stop generating speech immediately.}
    \label{fig:enter-label}
\end{figure}

\begin{figure}[h]
    \centering
    \includegraphics[width=\linewidth]{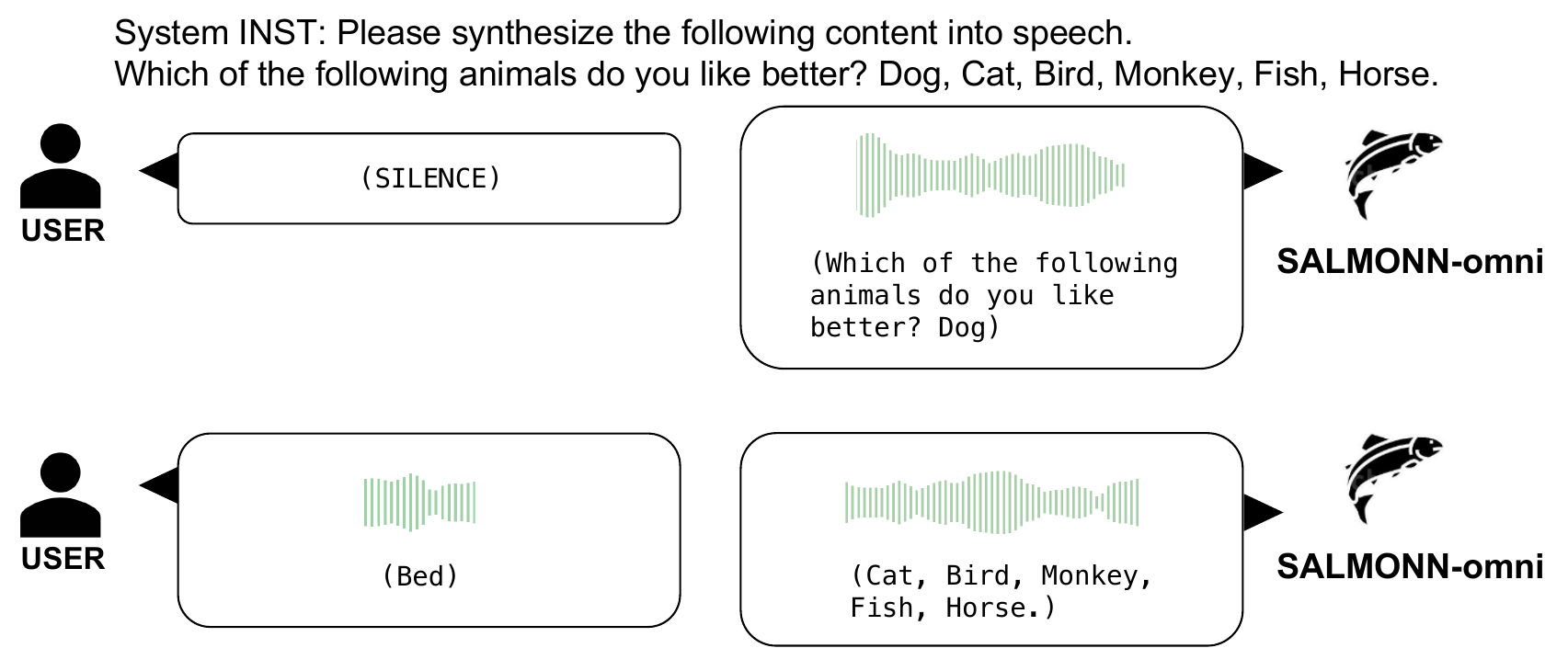}
    \caption{Context-dependent barge-in. When the user says something irrelevant, SALMONN-omni can ignore the input and continue speaking.}
    \label{fig:enter-label}
\end{figure}

\section{Conclusion}
This paper presents SALMONN-omni, a multimodal LLM developed within a novel codec-free, full-duplex framework for simultaneous speech understanding and generation. By integrating a streaming speech encoder, an LLM, and a streaming speech synthesizer into an end-to-end model, and incorporating a novel ``thinking'' strategy, SALMONN-omni can effectively unify a wide range of streaming speech tasks, including speech recognition, enhancement, dereverberation, target speaker extraction, and spoken question answering. Simulated experiments on turn-taking and context-dependent barge-in demonstrate SALMONN-omni's potential as a prototype of a future conversational AI that enables the seamless modelling of diverse interactive natural spoken dialogue dynamics using a single end-to-end model.

\newpage
{
\small
\printbibliography
}

\end{document}